\documentclass{sf2a-conf2015}
\usepackage{graphicx}
\usepackage{hyperref}
\usepackage[]{natbib}  
\usepackage{epstopdf}

\def\BibTeX{{\rm B\kern-.05em{\sc i\kern-.025em b}\kern-.08em
    T\kern-.1667em\lower.7ex\hbox{E}\kern-.125emX}}
\bibpunct{(}{)}{;}{a}{}{,}  
\newcommand{\kms}{{\mathrm{km~s^{-1}}}}

\newcommand{\caii}{\hbox{Ca$\;${\sc ii}}}
\newcommand{\hal}{\hbox{H${\alpha}$}}

\def\ms{\hbox{m\,s$^{-1}$}}
\def\kms{\hbox{km\,s$^{-1}$}}

\def\pstar{\hbox{$P_{\rm rot}$}}

\def\degr{\hbox{$^\circ$}}

\begin{document}

\TitreGlobal{SF2A 2015}


\title{SPIRou: a spectropolarimeter for the CFHT}

\runningtitle{SPIRou}

\author{C. Moutou}\address{Canada-France-Hawaii Telescope Corporation, 65-1238 Mamalahoa Hwy, Kamuela, HI 96743, USA}

\author{I. Boisse}\address{Universit\'e Aix-Marseille / CNRS-INSU, LAM / UMR 7326, 13388 Marseille, FRANCE}

\author{G. H\'ebrard$^{3,}$}\address{IAP / UMR 7095, CNRS / Universit\'e Pierre \& Marie Curie, 98 bis boulevard Arago, 75014 Paris, France} \address{Observatoire de Haute-Provence, CNRS, Universit\'e d'Aix-Marseille, 04870, Saint-Michel-l'Observatoire, France}

\author{E. H\'ebrard}\address{Universit\'e de Toulouse, UPS-OMP, IRAP, 14 avenue E. Belin, Toulouse, F-31400 France}\address{CNRS, IRAP / UMR 5277, Toulouse, 14 avenue E. Belin, F–31400 France}

\author{J.-F. Donati}\address{Universit\'e de Toulouse, UPS-OMP, IRAP, 14 avenue E. Belin, Toulouse, F-31400 France}\address{CNRS, IRAP / UMR 5277, Toulouse, 14 avenue E. Belin, F–31400 France}

\author{X. Delfosse}\address{Universit\'e Grenoble Alpes, IPAG, BP 53,
  F–38041 Grenoble Cedex 09, France}\address{CNRS, IPAG / UMR 5274, BP
  53, F–38041 Grenoble Cedex 09, France}

\author{D. Kouach}\address{Universit\'e de Toulouse, UPS-OMP, IRAP, 14 avenue E. Belin, Toulouse, F-31400 France}\address{CNRS, IRAP / UMR 5277, Toulouse, 14 avenue E. Belin, F–31400 France}

\author{the SPIRou team}

\setcounter{page}{237}


\maketitle


\begin{abstract}
SPIRou is a near-infrared spectropolarimeter and high-precision radial-velocity instrument, 
to be mounted on the 3.6m Canada-France-Hawaii telescope ontop
Maunakea and to be offered to the CFHT community from 2018. It focuses on two
main scientific objectives : (i) the search and study of Earth-like planets
around M dwarfs, especially in their habitable zone and (ii) the
study of stellar and planetary formation in the presence of stellar
magnetic field. The SPIRou characteristics (complete coverage of the
near infrared wavelengths, high resolution, high stability and
efficiency, polarimetry) also allow many other programs, e.g., magnetic
fields and atmospheres of M dwarfs and brown dwarfs, star-planet 
interactions, formation and characterization of massive stars, 
dynamics and atmospheric chemistry of planets in the solar system.
\end{abstract}



\section{Introduction}
SPIRou is a near-infrared spectropolarimeter designed for the
detection of exoplanets around low-mass stars and for the detection of
magnetic fields of young stellar objects. It is currently under
construction with a first light foreseen in fall 2017. SPIRou will
then be available to the community in 2018.  
An overview of the key aspects of SPIRou's
design is given in \citet{2014SPIE.9147E..15A}. \citet{2013sf2a.conf..497D} and 
\citet{2013sf2a.conf..509S} detailed several aspects of the science cases. In this article, we will recall the
science goals and challenges (Section 2), present the status of the instrument
development (Section 3) and
describe the SPIRou Legacy Survey (Section 4) as presently foreseen for the
Canada-France-Hawaii Telescope (CFHT).

\section{The SPIRou science}

\subsection{The Quest for Earth-like planets around M dwarfs}

	One objective of the current exoplanet science is to find and
        characterise habitable Earths and Super Earths. Among the 1600
        planets detected (and many more candidates), very few are in
        the habitable zone (sphere around a star where liquid water can be stable of
        the surface) of their host stars, and none is equivalent to Earth. 
	On another hand, very few planets are known around M
        dwarfs. Finally, observing the planetary architecture as a
        function of the stellar host mass is important to constrain planet formation models.
	
The long-term goal of exploring the atmospheric composition of the
telluric, potentially habitable exoplanets starts with the discovery
of such objects in the solar neighbourhood and the statistical description of their
properties. Ideally, one would need to complete a catalog of habitable
terrestrial planets, allowing future space missions to concentrate on
planet characterization without wasting time on their detection. In
addition, statistical properties of planets around M dwarfs can
provide key information on planetary formation, and in particular on
the sensitivity of planet formation to initial conditions in the
protoplanetary disc. Recent simulations indeed suggest that planet
populations change very significantly for M dwarfs of increasing
masses (Alibert et al in preparation). M dwarfs also vastly dominate
the stellar population in the Solar neighborhood and are likely
hosting most planets in our Galaxy; this makes such studies even more
crucial. Finally, pioneering studies with HARPS at the ESO 3.6m
telescope demonstrated that super-Earths with orbital periods <100 d
are more numerous around M dwarfs than around Sun-like stars, with an
occurrence frequency close to 90\% \citep{bonfils2013}; moreover,
about half of these super-Earths are potentially located in the
habitable zones of their host stars. 

	The exoplanetary scientists have several reasons to focus 
        on low-mass M dwarfs around which
        small/low-mass planets are easier to detect. Indeed, the radial-velocity signal of a planet
        is inversely proportional to $M_{\star}^{2/3}$, 
 so the method is more favorable to the
        coolest dwarf stars.
        It is also faster and easier to
        detect planets in the habitable zone of low-mass stars since it is
        located at a distance where planets orbital periods are in the
        range 10 to 50 days, thanks
        to a lower stellar temperature and to the fact that the radial-velocity
        semi-amplitude is proportional to $P_{orb}^{-1/3}$. 

Since M dwarfs are faint in the visible,
        we designed the SPIRou spectropolarimeter in the infrared, so
        that it is optimised
        for accurate radial-velocity measurements on M dwarfs.
        Two challenges will have yet to be faced when pursuing the search
        for telluric planets in the habitable zones of M dwarfs with SPIRou.
	First, numerous telluric lines are present in the
        near-infrared part of
        the spectra. These lines do not have an equal contribution
        on each observation depending on time and weather
        conditions. One should carefully mask or remove them, before
        deriving the radial-velocity of the star. 
	The second limitation is the level of activity of
        low-mass M dwarfs. We are currently investigating several
        reseach directions to mitigate this problem (see Sect. 2.3).
	
	The quest for low-mass planets around very low mass stars is a major science goal that requires large amounts of time. 
  
	\subsection{Characterization of transiting exoplanets}
Numerous ground- and space-based photometric surveys have been used to
detect new transiting planets and several others are in
project. Exoplanets transiting in front of their host stars are
particularly interesting objects as they allow numerous key studies to
be performed. This includes accurate radius, mass, and density
measurements, atmospheric studies in absorption through transits and
in emission through occultations, dynamic analyses from possible
timing variations or obliquity measurements. 

With precise masses of transiting exoplanets, the atmospheric models
can be efficiently tested. For instance, the presence of an atmosphere
and its composition strongly depend on these fundamental parameters,
as shown by the difference between CoRoT-7b and GJ 1214b (larger
radius for a similar mass for the latter): the atmosphere can be
dominated by water or by hydrogen, leading to different chemistry and
various densities. The diversity of mass or
density in the regime of Earth-like exoplanets and super-earthes is
striking, and requires a greater sample, and accurate mass
measurements, to become useful constraints for the theory
\citep{valencia2010, lopez2013, valencia2013}.

M dwarfs are advantageous hosts for transiting planets studies, as for
the radial-velocity method. Their
small radii make deeper the transits of telluric planets, and the
reduced size of their habitable zone makes more frequent and probable
the transits of a potentially habitable planet. SPIRou is thus well
optimized to perform the radial velocity follow up of photometric
surveys observing M dwarfs to detect transiting planets. Since the
rate of false positives could be significant, in particular those due
to blended or unblended binaries, the first goal of this follow up is
to establish the planetary nature of the transiting candidates. For
such validated transiting planets, the radial velocities are then used
to characterize the systems and in particular to measure the mass,
eccentricity, and obliquity of the planets. They are also used to
characterize the host stars and to search for possible extra
companions. SPIRou is expected to play an important role in the follow
up of planetary candidates transiting in front of M dwarfs detected by
photometric surveys, including ExTra, NGTS, K2, TESS, or PLATO.

Finally, as the magnetic field of the host stars will also be obtained
by SPIRou for each planet detection, it will be possible to
investigate the role of the magnetic field to shape the evolution of
planets, through evaporation or stellar wind interactions. Such studies
in transiting systems, where the relative inclinations are better
constrained and the planet masses are well known, are particularly interesting.

        \subsection{Filtering out the radial-velocity activity jitter using the spectropolarimetry}

To diagnose the radial velocity jitter, several complementary approaches are commonly used, mostly making use of chromospheric activity indicators like excess flux in the cores of the \hal\ and \caii\ H\&K, or measurements of spectral line asymmetries. 
For the use of SPIRou, we investigate how we can use
spectropolarimetry to better characterize and ultimately model the
radial-velocity jitter.
This new method aims at studying both the radial-velocity jitter caused by activity and Zeeman signatures reflecting the large-scale magnetic field at the origin of activity. 
To model the radial-velocity jitter due to rotational modulations, we take advantage of the distortions of the Stokes I profiles that are caused by brightness inhomogeneities at the stellar surface (H\'ebrard et al., 2015, in prep). 

\begin{figure}[ht!]
 \centering
\includegraphics[width=2.5in]{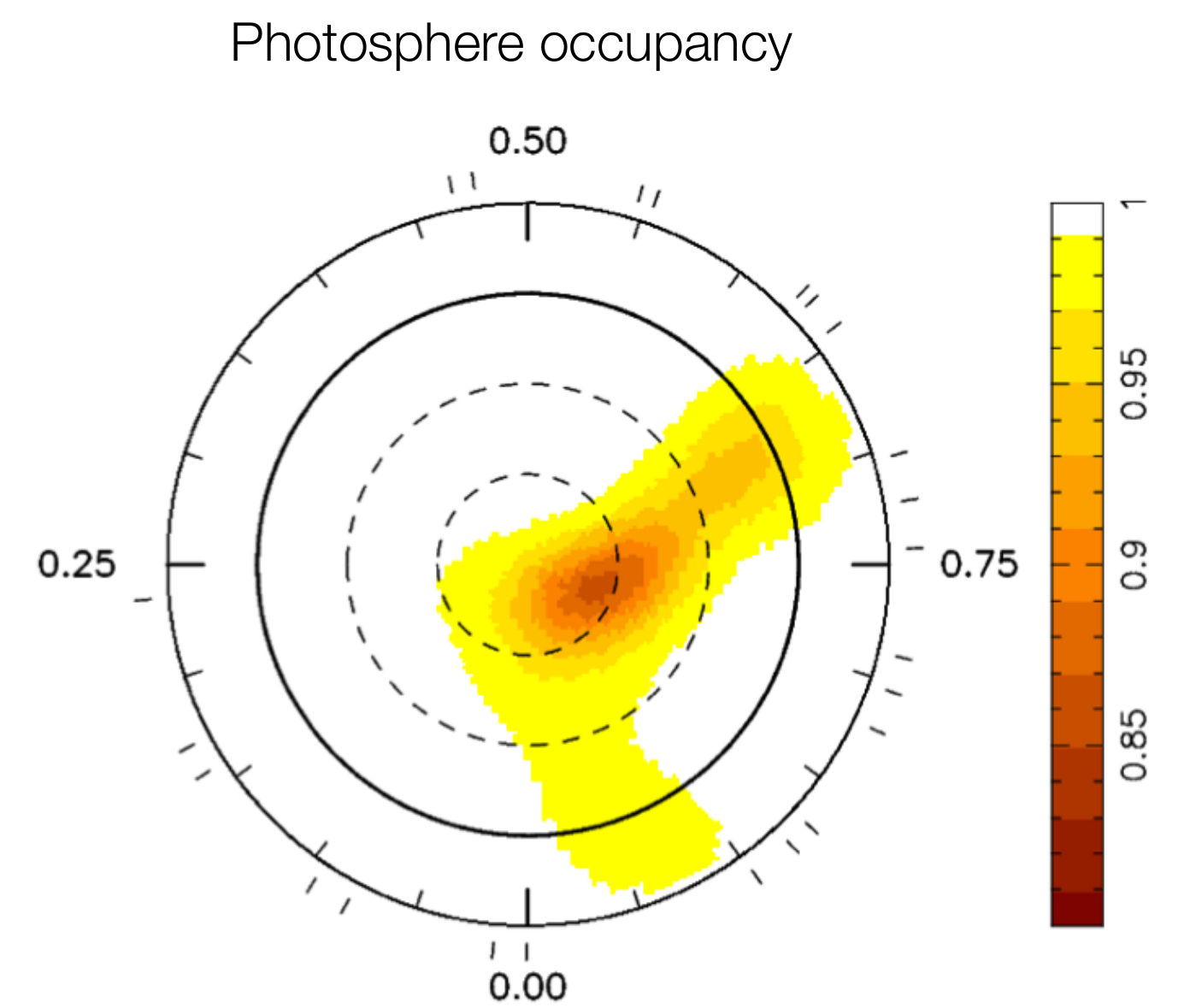}
\includegraphics[width=3in]{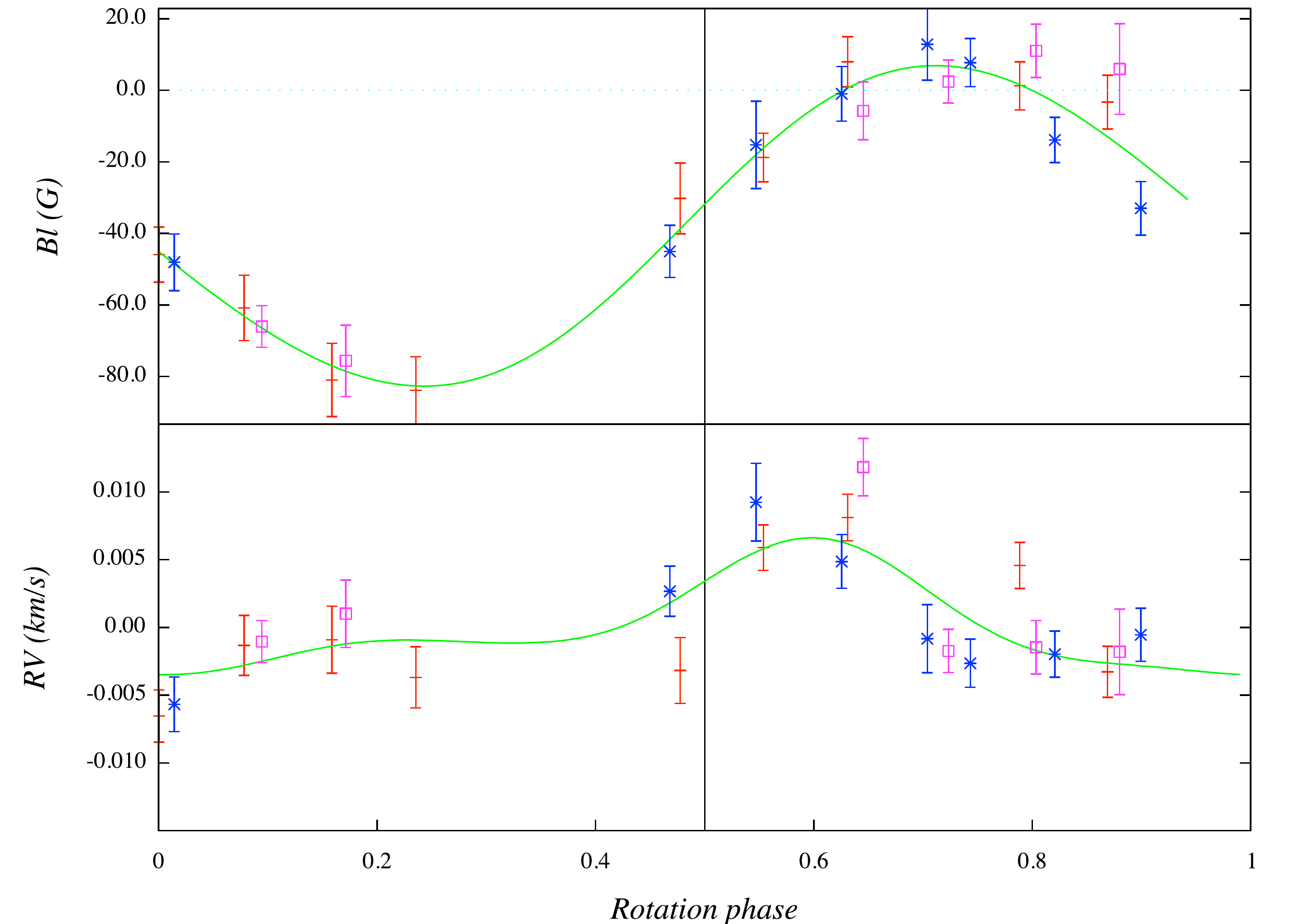}
\caption{\textit{Left:} Map of the photosphere occupancy  (white for quiet photosphere, brown for spot). The star is viewed in flattened polar projection. The pole is in the center, the bold circle depicts the stellar equator and the outer line represents  the -30\degr\ latitude. 
{\it Right:} Longitudinal field (in G) and radial velocities (in \kms)  as a function of the rotation phase, collected with HARPSPol 
Each color symbol represents a stellar rotation phase. The green curves represent the predicted values.
}
\label{fig:example2}
\end{figure}

As an example, we outline the results obtained for GJ~358, a moderately active M2-dwarf. From the Stokes $V$ signatures we infer a rotational period of $\sim$25.4 days; using Zeeman Doppler Imaging, we then recover the parent large scale magnetic field. 
This star exhibits a poloidal component with a polar intensity of -110~G  inclined at 45\degr to the line of sight.
We observe that the radial velocities and the longitudinal field vary in quadrature, which suggest that spots cluster at phase $\sim$0.7 (see right panel Fig.~\ref{fig:example2}). 
By modeling the tiny distortions in Stokes $I$ profile using a
modified version of Zeeman Doppler Imaging (H\'ebrard et al. 2015 in prep.), we obtain a
map representing the statistical distribution of spots at the stellar
surface. As expected, we find a cluster of spots lying close to the
same longitude. The radial velocity jitter of full amplitude of
$\sim$~15~\ms\ can be filtered out at a rms precision of 2.8 \ms\
close to the intrinsic precision of the data set (see right panel of
Fig.~\ref{fig:example2}). In the Lomb-Scargle periodogram of the
radial velocities, the periods \pstar, \pstar/2 and \pstar/3 have been filtered, and there is no more prevailing peaks in the periodogram of the residuals.\\
A similar method has been used the analysis of V830 Tau
\citep{donati2015} and it will be generalized to the planet search
programs in the SPIRou data. In the near-infrared, profile distortions
induced by cool spots are smaller than in the visible \citep{huelamo}, but the Zeeman
broadening is larger, so filtering needs to be adjusted accordingly \citep{hebrard2014}.

\subsection{Magnetic fields of protostars and cool stars}

Our second main scientific goal is to characterise the impact of
magnetic fields on the formation of stars and planets. 
Magnetic fields in the inner disc regions or at protostar surfaces can
only be investigated through nIR spectroscopy \citep{krull} or
optical spectropolarimetry \citep{donati05}. In both cases,
magnetic fields are detected through the Zeeman effect on spectral
lines, and more specifically through the additional broadening 
(with respect to non magnetically-sensitive lines of otherwise similar
properties) 
and the polarization structures they generate within line profiles. 
With SPIRou, we will measure
and characterise the structure of the magnetic field of young
protostars thanks to the polarimetric capability of SPIRou. The near
infrared is needed since these targets are faint in the
visible. Moreover the Zeeman effect increases with wavelength. 

The questions that SPIRou will be able to handle
include: (i) what is the origin of their magnetic fields (e.g. fossil
or dynamo), (ii) how do these magnetic fields connect the protostars
to their accretion discs, (iii) how do these fields control accretion
from the discs and how can they slow down the rotation of the
protostars, (iv) how do they participate (along with accretion disc
fields) in launching collimated jets, (v) how importantly do they
contribute to the photo-evaporation of the disk through the coronal
X-rays they indirectly produce, (vi) to what extent do magnetism and
magnetospheric accretion modify the internal structure of protostars,
and (vii) with which magnetic topologies do protostars start 
their unleashed spin-up towards the main sequence once they 
dissipated their accretion discs?

At ages of 0.1-0.5 Myr, low-mass pre-main-sequence stars
(class-I protostars) are surrounded by massive accretion discs from
which they actively feed, and are embedded in dust cocoons hiding them
from view at optical wavelengths. As a result, the large-scale
magnetic topology at the surface of a class-I protostars has never
been observed and will be assessed for the first time with SPIRou.

With SPIRou data, quantitative studies of how protostars accrete
material from their accretion discs and how this process impacts their
structure and rotation evolution can be carried out 
and the corresponding models can be directly confronted to
observations. One can for instance retrieve the location and geometry of
accretion funnels \citep{romanova}, investigate how the resulting accretion
shock affects the atmosphere of the star and evaluate the impact of
accretion on structure
and angular momentum content \citep{zanni} of the newly born star. 
In particular, class-I protostars should
be well suited for investigating magnetospheric accretion at higher
accretion-rate regimes.

With SPIRou, we will be estimating the gaseous content in the inner
regions of different types of protostellar accretion discs, investigating their magnetic field
strengths and topologies and
looking for potential correlations between the disc and field properties
to find out whether and how fields can impact discs.
Magnetic fields in discs are potentially important since fields are
expected to impact significantly the formation, disc-induced migration
and orbital inclination of protoplanets. More specifically, discs
fields and associated MHD turbulence can potentially inhibit
fragmentation through gravitational instabilities and modify the
subsequent formation of giant planets. Thanks to our first study on
filtering out 
activity jitter on rotating stars \citep{petit2015,donati2015}, we
could also search for hot Jupiters around these
targets. Detecting hot Jupiters at the early stage would constrain the
theory on the formation and migration of planetary systems.

\section{The instrument development status}
SPIRou is being developed by seven countries: Canada, France,
Switzerland, Taiwan, Brazil, Portugal, Hawaii and 12 institutes in
these countries. The
consortium is led by IRAP (Toulouse, France).

The science requirements and technical specifications of SPIRou can be
summarized as follows: 
\begin{itemize}
\item Acquire a single-shot spectral domain covering the YJHK domain (0.98-2.35
  $\mu$m), to benefit the very high spectral radial velocity
  content in the H and K-band for M-dwarfs and to allow observing the
magnetic field of embedded young stars in their early phase.
\item Get a spectral resolution of 70K to 75K, in order to 
optimize the radial-velocity accuracy for lines of slowly rotating M dwarfs.
\item Achieve the radial velocity stability of 1m/s, to allow the
detection of Earth and Super-Earth exoplanets in the habitable zone of M-dwarfs.
\item Provide the linear and circular polarimetric capacity, with maximum
1-2 \% crosstalk and down to a sensitivity of 10 ppm (similar to
CFHT/ESPaDOnS). This will allow the observations of stellar magnetic fields
and the disentangling of the stellar activity and the keplerian signal. 
\item Obtain a signal-to-noise ratio of~100 per 2km/s pixel for
  a star of magnitude J=12.0 or K=11; this implies a requirement on the
throughput of 15 \% and on the thermal background at 2.35 $\mu$m
of maximum 50 ph/s/\AA.  This efficiency is required for a large
radial-velocity and polarimetric survey.
\end{itemize}

\begin{figure}[ht!]
 \centering
\includegraphics[width=5in]{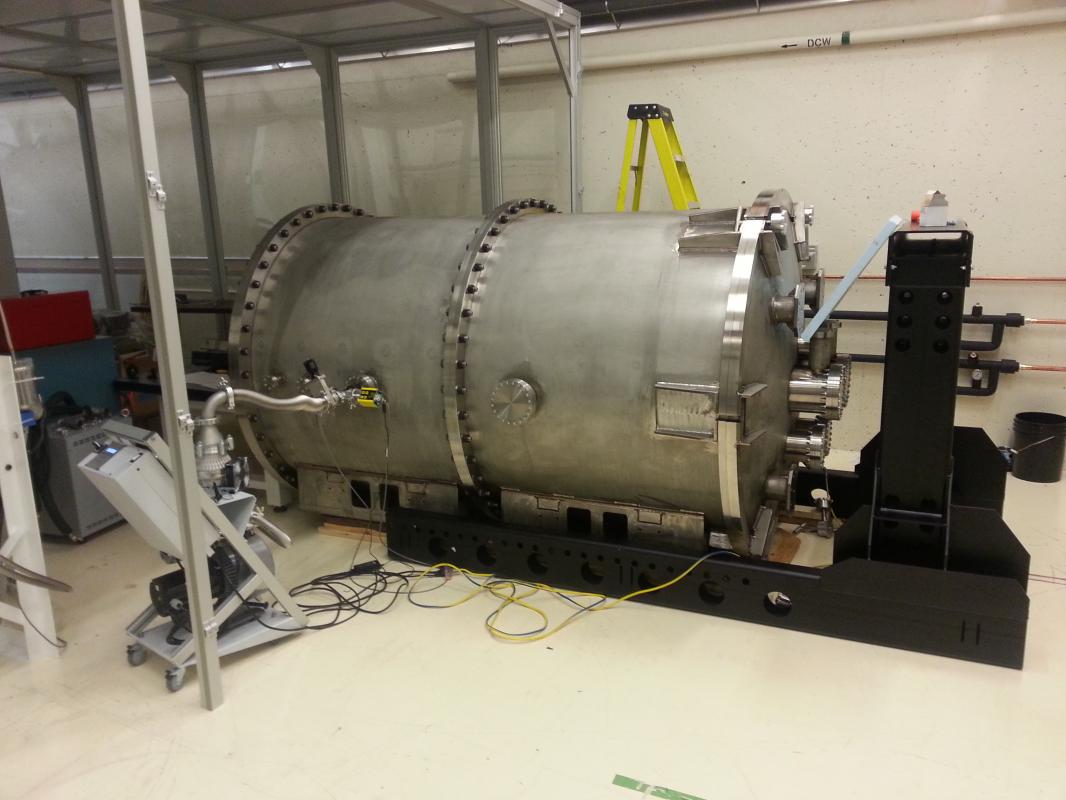}
\caption{The SPIRou cryostat, being integrated at NRC Herzberg,
  Victoria, Canada.}
\label{fig:example3}
\end{figure}

SPIRou is made of several sub-systems, each of them being
simultaneously developed in different institutes of the
consortium. The first sub-system is the Cassegrain unit, mounted in
the dome at the Cassegrain focus of the 3.6m CFH telescope. It is composed of an
atmospheric dispersion corrector, the guide camera with an 
image stabilisation system and an achromatic polarimeter. The
Cassegrain unit is being developed and tested at IRAP, Toulouse. The second
sub-system contains the fibers and the pupil slicer. The spectral
range requires the use of fluoride fibers; high transmission fibers
are chosen to maximize thoughput. The
near-field scrambling is obtained with a section of octogonal
fibers. The pupil is split into four slices for each science fiber,
with 4 pixels separating each fiber. The main piece of SPIRou is the
cryogenic spectrograph, located in the coud\'e room at the third floor
of the telescope building. Its thermal control has a specification of 2
mK over a day. The spectrograph design contains a large
parabolic mirror, a double-pass prism-train cross-disperser with an R2
diffraction grating, and a fully dioptric camera with a Hawaii 4RG 4kx4k
detector array. The spectrograph cryostat, currently being mounted in
Victoria (Canada),
is shown on Fig.~\ref{fig:example3}.
Essential to the radial-velocity accuracy is the
calibration \& super-stable radial-velocity reference module, also
located in the coud\'e room and fiber linked to the spectrograph; this
calibration module is being assembled between Geneva and Observatoire
de Haute Provence.

The data reduction pipeline is being developed by the consortium
(effort led by LAM, Marseille) and will be delivered to CFHT with the
instrument, aiming at a complete extraction of spectra as well as
radial-velocity and polarisation signatures available a few minutes after the observations.

The last project review was held in June 2015 at IRAP, Toulouse. The
funding process is now finalised, and should be secured at the end of
2015. The integration / validation of the full instrument will be at
IRAP, Toulouse and should last 7 months. The integration at system
level is expected to start on March 2016, followed by the validation
phase on November 2016. The system Acceptance at IRAP will be in April
2017. We expect the final integration on CFHT site to be in summer
2017 for a first light in the last semester of 2017.

Stay tuned by consulting the SPIRou web page \url{http://spirou.irap.omp.eu/}.

\section{The SPIRou Legacy Survey (SLS)}
Shortly after its installation on the CFHT, SPIRou will start a series
of Legacy Surveys, for a minimum of 300 nights over the first 3 years of
operations of the
instrument -as approved by the CFHT Board in 2015. The objectives of the surveys are threefold:
\begin{itemize}
\item identify several tens of habitable-zone Earth-like planets orbiting M
  dwarfs in the
  immediate neighborhood of the Solar System, as well as more than 100
  exoplanets with orbital periods ranging from 1 day to 1 year, to derive accurate planet statistics,
\item characterize the internal structure of several dozens of
  telluric planets with moderate equilibrium temperature, by providing
  mass measurements to the transit candidates discovered by K2, TESS
  and PLATO,
\item explore the impact of magnetic fields on star / planet formation, by detecting and characterizing magnetic fields of various types of young stellar objects.
\end{itemize}

These three observing programs require large input catalogs of
targets for statistical relevance, and a large number of visits per
target. With its queue operation mode, the CFHT is well adapted and
flexible to handle large surveys, get homogeneous data sets and
optimize temporal sampling and critical scheduling. Specific tools are
being developed at the observatory to support SPIRou surveys
operations.\\

The radial-velocity {\bf planet-search survey (SLS-PS)} will focus on an input catalog of
330 nearby M dwarfs. They are selected from the catalog by
\citet{lepine} and cover a range in mass from 0.08 to 0.5 M$_{sun}$.
Monte Carlo simulations suggest that the SLS-PS should detect $\sim250$ new
exoplanets, including $\sim$180 with masses less than 5 Earth masses and
$\sim$30 of them lying in the habitable zone. Breaking down the full stellar sample in
0.5-dex bins in orbital period and mass, the SLS-PS will provide
occurrence rates of Earth-like planets in each mass / period bin with
a $\sim$20\% accuracy. In particular, the SLS-PS will determine the
frequency $\eta_E$ of habitable-zone Earth-like planets around M dwarfs to an accuracy of $\sim$5\%.
Extensions of this program are possible and can be carried out after
the completion of SLS-PS. The required number of visits per star
averages to $\sim$ 50 per star during the survey duration, but this
parameter may have to be increased at the expanse of decreasing the
sample size, especially if multi-planet systems
are frequently found or/and if the stellar jitter remains an issue.

The total time required for the SLS-PS is of the order of 275 nights,
counting on $\sim$ 10 min per visits to achieve the required 1 m/s
precision on individual radial velocity measurements. As SPIRou observations are
done in the spectropolarimetric mode, the magnetic field of the host
stars will be monitored at the same frequency than the planetary
orbits, offering a unique view on the evolution of magnetic topologies
of evolved and quiet fully convective stars. Other activity proxies
will also be monitored.

With the future collection of planetary systems to be found by SPIRou,
we will be able to bring unprecedented constraints on the formation of
planets around low-mass stars, their dynamical evolution and
interactions between planets and with the host star.
Consequences for further studies are expected to be huge, as
tidal locking, orbital properties and stellar wind may be limits to
planet habitability and presently require additional observational
constraints.\\

The radial-velocity follow-up observations of transiting planetary candidates is
the best way to get rid of false positives \citep{santerne} and to
measure the mass of the confirmed planets. With mass and radius, the
bulk density and global properties of the planetary internal structure
can be inferred. In the continuation of successful missions like
CoRoT and $Kepler$, NASA is currently operating the K2 mission 
and is going to launch TESS in 2017. By targeting bright stars, TESS
will provide Earth-like planet candidates which radial-velocity follow-up will be
possible on ground-based telescopes -unlike most of the $Kepler$ candidates.
SPIRou at CFHT will be unique to secure radial-velocity measurements on the coolest
part of the stellar hosts. Using statistics based on Kepler results,
TESS predicts the discovery of several hundreds of Earth-like and
super-Earth planets, in addition to thousands of icy and gas giants,
for stars brighter than I=12. The list of planet candidates discovered
by TESS and their light curves will be made available to the community
for complementary observations. Most super-Earth candidates detected by TESS
 will orbit M dwarfs - with less than ~30\% of them
accessible to optical radial-velocity follow-ups; nIR velocimeters like SPIRou will
thus be essential to validate planet candidates. 

In the meantime, ground-based transit surveys of M stars are ongoing
(e.g., MEarth) or about to start (e.g., ExTrA), with many candidate
planets known by 2017. With a first light scheduled in 2016, ExTrA
targets stars of late spectral types (M4 to L0) and aims at the
smallest planets around the smallest stars.  ExTrA should detect a few
tens of very-low-mass transiting planet candidates, again to be
validated with nIR velocimetry.
Expected to even outperform ESPRESSO on the VLT for most types of M
dwarfs, SPIRou will be the best and most efficient radial-velocity instrument to
monitor in the nIR the candidates visible from CFHT, to confirm 
 or reject their planetary nature and to determine their masses.

The second component of the SLS - {\bf the SLS-TF for Transit Follow-up},
will carry out radial-velocity follow-up observations of the 50 most interesting
transiting planet candidates uncovered by future photometry surveys
including TESS and ExTrA - for a total of 100 CFHT nights. We will
focus on candidates with orbits closest to their habitable zones and/or around the
brightest stars for a given spectral type - to select the best ones
to be scrutinized in the future by the JWST or the E-ELT for spectral
signatures of biomarkers in their atmospheres. The observing strategy
includes an estimated number of 70 visits per candidate. 
 \\

The third component of the SPIRou Legacy Survey is the study of
class-I,-II and -III protostars ({\bf SLS-MP: Magnetic
Protostar / Planet}). We aim at detecting and mapping the magnetic topologies of
young stars and their accretion discs and to explore the origins and
evolution of the magnetic field of stars. In this survey, it will also be
possible to unveil giant planets in short orbits and put constraints
on the formation versus migration timescales for hot-Jupiter like
planets - as they survive the system formation processes.

Magnetic fields play a major role in the early evolution of stellar
systems. By affecting and often controlling accretion processes,
triggering and channeling outflows, and producing intense X-rays,
magnetic fields critically impact the physics and largely dictate the
angular momentum evolution of pre-main-sequence stars and their accretion
discs.  Magnetic fields link accreting pre-main-sequence stars to their discs and
evacuate the central disc regions; they funnel the disc material onto
the stars, generate powerful winds 
and force classical T Tauri stars to rotate much more slowly than expected from
contraction and accretion. Fields of accretion discs trigger
instabilities, enhance accretion rates and extract angular momentum through magnetic
braking and outflows; they also produce X-rays through flares and
shocks, ionizing the disc gas and impacting disc dynamics and planet
formation.

The targets are brighter in the nIR than in the optical, whereas the
Zeeman effect is stronger at longer wavelengths. Spectropolarimetric
observations with SPIRou will thus allow us to survey much larger
samples of pre-main-sequence stars than accessible presently, as well as to access
younger embedded class-I protostars and their discs on which very
little information is yet available.

The SLS-MP survey will consist in collecting spectropolarimetric observations of
about 140 young protostars in various stages of evolution, selected in
the close star-forming regions of Taurus/Auriga, TW Hya and $\rho$
Ophiuchius. It will, in particular, explore for the first time the
magnetic topologies of class-I protostars. 
The survey will extend over 125 nights as each target
requires multiple visits (about 20) to sample the rotation cycle over
a couple of cycles. As previously, the MP survey can be extended
towards other young systems. \\

With these three axes, the SPIRou Legacy Survey will gather a large
community of scientists interested in the formation, evolution and
characteristics of exoplanets and stars. The SPIRou science group is
open to any proposal of legacy-type science that would not yet be
covered but accessible with the survey observations. Out of these surveys, SPIRou
will be able and available to explore many more science topics via
open time proposals. 

\section{Conclusions} 

The SPIRou near-infrared velocimeter and spectropolarimeter, soon to
be installed at the Maunakea CFHT observatory, mainly aims at the 
specific tasks of investigating planet formation and properties around
cool stars and young stars, while describing the magnetic properties
of these stars. The polarimetric capability is unique and will widen
the future diagnostics of stellar variability for planet
search. SPIRou has the potential of finding the telluric exoplanets
that space telescopes like JWST will focus on, for atmospheric studies
and further habitability explorations. Other near-infrared
velocimeters are being built in the world, e.g. CARMENES
\citep{quirrenbach}, IRD \citep{kotani}, HPF \citep{hearty}. None of
these instruments, however, includes the spectropolarimetric
capability (required for stellar activity corrections) nor the
essential K band (with a large radial-velocity content for low mass
stars and a large flux for embedded sources).

The SPIRou science group is open to any interested member of the CFHT
community, especially in the context of the large surveys which will
use SPIRou for $\sim$ 70\% of the time and at least 300 nights over
the 3 first years.


 The community is now structured around five main work packages; three
 focussing on the three components of the SPIRou Legacy Survey.  The
 other two explore topics that are common to
 all science goals, like radial-velocity optimization, stellar jitter
 correction or stellar spectroscopic and magnetic characterisation. 

The first sub-systems of SPIRou are being delivered to Toulouse, 
and the integration and tests will take place at IRAP until mid 2017,
when the instrument will be shipped to Hawaii. SPIRou should be on the sky
at the end of 2017, and probably open to the community in 2018A.




\bibliographystyle{aa}
\bibliography{moutou2} 

\begin{thebibliography}{20}
\expandafter\ifx\csname natexlab\endcsname\relax\def\natexlab#1{#1}\fi

\bibitem[{{Artigau} {et~al.}(2014){Artigau}, {Kouach}, {Donati}, {Doyon},
  {Delfosse}, {Baratchart}, {Lacombe}, {Moutou}, {Rabou}, {Par{\`e}s},
  {Micheau}, {Thibault}, {Reshetov}, {Dubois}, {Hernandez}, {Vall{\'e}e},
  {Wang}, {Dolon}, {Pepe}, {Bouchy}, {Striebig}, {H{\'e}nault}, {Loop},
  {Saddlemyer}, {Barrick}, {Vermeulen}, {Dupieux}, {H{\'e}brard}, {Boisse},
  {Martioli}, {Alencar}, {do Nascimento}, \& {Figueira}}]{2014SPIE.9147E..15A}
{Artigau}, {\'E}., {Kouach}, D., {Donati}, J.-F., {et~al.} 2014, in Society of
  Photo-Optical Instrumentation Engineers (SPIE) Conference Series, Vol. 9147,
  Society of Photo-Optical Instrumentation Engineers (SPIE) Conference Series,
  15

\bibitem[{{Bonfils} {et~al.}(2013){Bonfils}, {Delfosse}, {Udry}, {Forveille},
  {Mayor}, {Perrier}, {Bouchy}, {Gillon}, {Lovis}, {Pepe}, {Queloz}, {Santos},
  {S{\'e}gransan}, \& {Bertaux}}]{bonfils2013}
{Bonfils}, X., {Delfosse}, X., {Udry}, S., {et~al.} 2013, \aap, 549, A109

\bibitem[{{Delfosse} {et~al.}(2013){Delfosse}, {Donati}, {Kouach},
  {H{\'e}brard}, {Doyon}, {Artigau}, {Bouchy}, {Boisse}, {Brun}, {Hennebelle},
  {Widemann}, {Bouvier}, {Bonfils}, {Morin}, {Moutou}, {Pepe}, {Udry}, {do
  Nascimento}, {Alencar}, {Castilho}, {Martioli}, {Wang}, {Figueira}, \&
  {Santos}}]{2013sf2a.conf..497D}
{Delfosse}, X., {Donati}, J.-F., {Kouach}, D., {et~al.} 2013, in SF2A-2013:
  Proceedings of the Annual meeting of the French Society of Astronomy and
  Astrophysics, ed. L.~{Cambresy}, F.~{Martins}, E.~{Nuss}, \& A.~{Palacios},
  497--508

\bibitem[{{Donati} {et~al.}(2015){Donati}, {H{\'e}brard}, {Hussain}, {Moutou},
  {Malo}, {Grankin}, {Vidotto}, {Alencar}, {Gregory}, {Jardine}, {Herczeg},
  {Morin}, {Fares}, {M{\'e}nard}, {Bouvier}, {Delfosse}, {Doyon}, {Takami},
  {Figueira}, {Petit}, {Boisse}, \& {MaTYSSE Collaboration}}]{donati2015}
{Donati}, J.-F., {H{\'e}brard}, E., {Hussain}, G.~A.~J., {et~al.} 2015, \mnras,
  453, 3706

\bibitem[{{Donati} {et~al.}(2005){Donati}, {Paletou}, {Bouvier}, \&
  {Ferreira}}]{donati05}
{Donati}, J.-F., {Paletou}, F., {Bouvier}, J., \& {Ferreira}, J. 2005, \nat,
  438, 466

\bibitem[{{Hearty} {et~al.}(2014){Hearty}, {Levi}, {Nelson}, {Mahadevan},
  {Burton}, {Ramsey}, {Bender}, {Terrien}, {Halverson}, {Robertson}, {Roy},
  {Blank}, {Blanchard}, \& {Stefansson}}]{hearty}
{Hearty}, F., {Levi}, E., {Nelson}, M., {et~al.} 2014, in Society of
  Photo-Optical Instrumentation Engineers (SPIE) Conference Series, Vol. 9147,
  Society of Photo-Optical Instrumentation Engineers (SPIE) Conference Series,
  52

\bibitem[{{H{\'e}brard} {et~al.}(2014){H{\'e}brard}, {Donati}, {Delfosse},
  {Morin}, {Boisse}, {Moutou}, \& {H{\'e}brard}}]{hebrard2014}
{H{\'e}brard}, {\'E}.~M., {Donati}, J.-F., {Delfosse}, X., {et~al.} 2014,
  \mnras, 443, 2599

\bibitem[{{Hu{\'e}lamo} {et~al.}(2008){Hu{\'e}lamo}, {Figueira}, {Bonfils},
  {Santos}, {Pepe}, {Gillon}, {Azevedo}, {Barman}, {Fern{\'a}ndez}, {di Folco},
  {Guenther}, {Lovis}, {Melo}, {Queloz}, \& {Udry}}]{huelamo}
{Hu{\'e}lamo}, N., {Figueira}, P., {Bonfils}, X., {et~al.} 2008, \aap, 489, L9

\bibitem[{{Johns-Krull}(2007)}]{krull}
{Johns-Krull}, C.~M. 2007, \apj, 664, 975

\bibitem[{{Kotani} {et~al.}(2014){Kotani}, {Tamura}, {Suto}, {Nishikawa},
  {Sato}, {Aoki}, {Usuda}, {Kurokawa}, {Kashiwagi}, {Nishiyama}, {Ikeda},
  {Hall}, {Hodapp}, {Hashimoto}, {Morino}, {Okuyama}, {Tanaka}, {Suzuki},
  {Inoue}, {Kwon}, {Suenaga}, {Oh}, {Baba}, {Narita}, {Kokubo}, {Hayano},
  {Izumiura}, {Kambe}, {Kudo}, {Kusakabe}, {Ikoma}, {Hori}, {Omiya}, {Genda},
  {Fukui}, {Fujii}, {Guyon}, {Harakawa}, {Hayashi}, {Hidai}, {Hirano},
  {Kuzuhara}, {Machida}, {Matsuo}, {Nagata}, {Onuki}, {Ogihara}, {Takami},
  {Takato}, {Takahashi}, {Tachinami}, {Terada}, {Kawahara}, \&
  {Yamamuro}}]{kotani}
{Kotani}, T., {Tamura}, M., {Suto}, H., {et~al.} 2014, in Society of
  Photo-Optical Instrumentation Engineers (SPIE) Conference Series, Vol. 9147,
  Society of Photo-Optical Instrumentation Engineers (SPIE) Conference Series,
  14

\bibitem[{{L{\'e}pine} \& {Gaidos}(2011)}]{lepine}
{L{\'e}pine}, S. \& {Gaidos}, E. 2011, \aj, 142, 138

\bibitem[{{Lopez} \& {Fortney}(2014)}]{lopez2013}
{Lopez}, E.~D. \& {Fortney}, J.~J. 2014, \apj, 792, 1

\bibitem[{{Petit} {et~al.}(2015){Petit}, {Donati}, {H{\'e}brard}, {Morin},
  {Folsom}, {B{\"o}hm}, {Boisse}, {Borgniet}, {Bouvier}, {Delfosse}, {Hussain},
  {Jeffers}, {Marsden}, \& {Barnes}}]{petit2015}
{Petit}, P., {Donati}, J.-F., {H{\'e}brard}, E., {et~al.} 2015, ArXiv e-prints

\bibitem[{{Quirrenbach} {et~al.}(2010){Quirrenbach}, {Amado}, {Mandel},
  {Caballero}, {Mundt}, {Ribas}, {Reiners}, {Abril}, {Aceituno}, {Afonso},
  {Barrado y Navascues}, {Bean}, {B{\'e}jar}, {Becerril}, {B{\"o}hm},
  {C{\'a}rdenas}, {Claret}, {Colom{\'e}}, {Costillo}, {Dreizler},
  {Fern{\'a}ndez}, {Francisco}, {Galad{\'{\i}}}, {Garrido}, {Gonz{\'a}lez
  Hern{\'a}ndez}, {Gu{\`a}rdia}, {Guenther}, {Guti{\'e}rrez-Soto}, {Joergens},
  {Hatzes}, {Helmling}, {Henning}, {Herrero}, {K{\"u}rster}, {Laun}, {Lenzen},
  {Mall}, {Martin}, {Mart{\'{\i}}n-Ruiz}, {Mirabet}, {Montes}, {Morales},
  {Morales Mu{\~n}oz}, {Moya}, {Naranjo}, {Rabaza}, {Ram{\'o}n}, {Rebolo},
  {Reffert}, {Rodler}, {Rodr{\'{\i}}guez}, {Rodr{\'{\i}}guez Trinidad},
  {Rohloff}, {S{\'a}nchez Carrasco}, {Schmidt}, {Seifert}, {Setiawan},
  {Solano}, {Stahl}, {Storz}, {Su{\'a}rez}, {Thiele}, {Wagner}, {Wiedemann},
  {Zapatero Osorio}, {del Burgo}, {S{\'a}nchez-Blanco}, \& {Xu}}]{quirrenbach}
{Quirrenbach}, A., {Amado}, P.~J., {Mandel}, H., {et~al.} 2010, in Society of
  Photo-Optical Instrumentation Engineers (SPIE) Conference Series, Vol. 7735,
  Society of Photo-Optical Instrumentation Engineers (SPIE) Conference Series,
  13

\bibitem[{{Romanova} {et~al.}(2011){Romanova}, {Long}, {Lamb}, {Kulkarni}, \&
  {Donati}}]{romanova}
{Romanova}, M.~M., {Long}, M., {Lamb}, F.~K., {Kulkarni}, A.~K., \& {Donati},
  J.-F. 2011, \mnras, 411, 915

\bibitem[{{Santerne} {et~al.}(2013{\natexlab{a}}){Santerne}, {Donati}, {Doyon},
  {Delfosse}, {Artigau}, {Boisse}, {Bonfils}, {Bouchy}, {H{\'e}brard},
  {Moutou}, \& {Udry}}]{2013sf2a.conf..509S}
{Santerne}, A., {Donati}, J.-F., {Doyon}, R., {et~al.} 2013{\natexlab{a}}, in
  SF2A-2013: Proceedings of the Annual meeting of the French Society of
  Astronomy and Astrophysics, ed. L.~{Cambresy}, F.~{Martins}, E.~{Nuss}, \&
  A.~{Palacios}, 509--514

\bibitem[{{Santerne} {et~al.}(2013{\natexlab{b}}){Santerne}, {Donati}, {Doyon},
  {Delfosse}, {Artigau}, {Boisse}, {Bonfils}, {Bouchy}, {H{\'e}brard},
  {Moutou}, \& {Udry}}]{santerne}
{Santerne}, A., {Donati}, J.-F., {Doyon}, R., {et~al.} 2013{\natexlab{b}}, in
  SF2A-2013: Proceedings of the Annual meeting of the French Society of
  Astronomy and Astrophysics, ed. L.~{Cambresy}, F.~{Martins}, E.~{Nuss}, \&
  A.~{Palacios}, 509--514

\bibitem[{{Valencia} {et~al.}(2013){Valencia}, {Guillot}, {Parmentier}, \&
  {Freedman}}]{valencia2013}
{Valencia}, D., {Guillot}, T., {Parmentier}, V., \& {Freedman}, R.~S. 2013,
  \apj, 775, 10

\bibitem[{{Valencia} {et~al.}(2010){Valencia}, {Ikoma}, {Guillot}, \&
  {Nettelmann}}]{valencia2010}
{Valencia}, D., {Ikoma}, M., {Guillot}, T., \& {Nettelmann}, N. 2010, \aap,
  516, A20

\bibitem[{{Zanni} \& {Ferreira}(2013)}]{zanni}
{Zanni}, C. \& {Ferreira}, J. 2013, \aap, 550, A99

\end{thebibliography}


%
\end{document}